\documentclass[12pt,nohyper,notoc]{JHEP}

\usepackage{amsmath,amssymb,cite} 
	
\newcommand{\aD}{{\dot\alpha}}

\newcommand{\bD}{{\dot\beta}}

\newcommand{\BL}{{\bf L}} 

\def\com{X}
\def\sfc{\hat\Omega}
\def\Z{{\mathbb Z}}

\def\N{{\cal N}}
\def\M{{\cal M}}
\def\sst{\scriptscriptstyle}
\def\SU{\text{SU}}
\def\SO{\text{SO}}
\def\Sp{\text{Sp}}
\def\O{\text{O}}
\def\U{\text{U}}
\def\K{{\cal K}}

\title{Instantons, finite N=2 Sp(N) theories and the AdS/CFT correspondence}

\author{Timothy J.~Hollowood\\
Theoretical Division T-8, Los Alamos National Laboratory,
Los Alamos, NM 87545, USA\\
Department of Physics, University of Wales Swansea,
Swansea, SA2 8PP, UK\\
E-mail: {\tt pyth@skye.lanl.gov}}

\abstract{We examine ADHM multi-instantons in the conformal $\N=2$
supersymmetric $\Sp(N)$ gauge theory with one 
anti-symmetric tensor and four fundamental hypermultiplets.
We argue that the ADHM construction and measure can also be deduced
from purely field theoretic considerations and also from 
the dynamics of D-instantons in the presence of D3-branes,
D7-branes and an orientifold O7-plane. The measure then admits a large-$N$
saddle-point approximation where the D3-branes disappear but the
background is changed to $AdS_5\times S^5/{\mathbb Z}_2$, as expected on the
basis of the AdS/CFT correspondence. The large-$N$ measure displays
the fractionation of D-instantons at the singularity $S^3\subset
S^5/{\mathbb Z}_2$ and is described for instanton number $k$ 
by a certain $\O(k)$ matrix model.} 

\keywords{Solitons Monopoles and Instantons, $1/N$ Expansion, Duality
in Gauge Field Theories, Supersymmetry and Duality}

\preprint{{\tt hep-th/9908201}}

\begin{document}

\section{Introduction}

In this paper, we shall be concerned with instanton
effects in the gauge theory which, in the context of $\N=4$
supersymmetric $\SU(N)$ gauge theory, have a very precise relation to
D-instanton effects in the string theory
\cite{BGKR,lett,MO-III,Green:rev} in perfect accord with the AdS/CFT
correspondence \cite{MAL,AGMOO:rev}. 
This relation can be seen in the dynamics of D-instantons and D3-branes in type IIB string
theory. The world volume theory of $N$ D3-branes is precisely 
the $\N=4$ supersymmetric $\U(N)$ gauge theory, in the low energy
decoupling limit where all energy scales are much smaller than
$(\alpha')^{-1/2}$ \cite{MO-III}. Instantons solutions in the gauge theory are identified
as D-instantons (or D$(-1)$-branes) bound to the D3-branes 
\cite{Witten:1996gx,Douglas:1995bn,Douglas:1996uz,MO-III}. The
D-instantons are described by a `world volume' $\U(k)$ matrix model,
for $k$ instantons, whose degrees-of-freedom correspond to the ADHM construction of
instantons. This theory is the dimensional reduction of
six-dimensional $\N=(1,0)$ supersymmetric gauge theory and has $\U(k)$
adjoint and fundamental degrees-of-freedom, the latter arising from
strings stretched between the D-instantons and D3-branes. In the
decoupling limit this ADHM matrix model is strongly coupled and only
certain terms in the action survive. The resulting partition function
is identical to the measure on the ADHM moduli space weighted by the
instanton action constructed by purely field theoretical methods in
\cite{meas1,meas2,MO-III}. Remarkably, auxiliary scalars introduced
to bi-linearize a certain four fermionic collective coordinate
interaction are interpreted as the freedom for the D-instantons to be
ejected from the D3-branes: they need not be bound!

At large $N$, the ADHM matrix model can be approximated.
The idea is to first integrate out the variables
corresponding to D$(-1)$-D3 strings, the resulting measure is then 
amenable to a standard saddle-point analysis. From the point-of-view of the
D-instantons the D3-branes disappear to leave a description of
D-instantons in a $AdS_5\times S^5$ background. Instantons are
therefore a perfect way to see the AdS/CFT correspondence in action.
It may seem surprising that the gauge theory instanton calculation is
done at large $N$ with $g^2N\ll1$ and yet we find perfect agreement
with the supergravity regime with $g^2N\gg1$. Apparently there is a
non-renormalization theorem at work, which is natural from the
supergravity side of the description \cite{Gopakumar:1999xh}, but has yet to be proved
from the gauge theory side. 

Since instantons prove such a powerful tool it is natural to study them
in other gauge theories which may or may not have a known dual
description. This has been undertaken in the following conformal field
theories:

(i) $\N=4$ supersymmetric gauge theory with gauge groups $\O(N)$ and
$\Sp(N)$ \cite{spo}. In these theories the dual is known to be the
type IIB superstring on $AdS_5\times{\mathbb R}P^5$ 
\cite{Kakushadze:1998tr,Kakushadze:1998tz}.

(ii) $\N=2$ supersymmetric $\SU(N)$ gauge theory with $2N$ fundamental
hypermultiplets \cite{N=2}. In this case the dual string theory is not known.

(iii) `Orbifold' theories which correspond to certain projections of
the $\N=4$ supersymmetric $\SU(N)$ theory \cite{orbi} by a finite
subgroup $\Gamma$ of the $\SU(4)$ $R$-symmetry group of the theory. 
The resulting theories have
either $\N=0,1$ or 2 supersymmetries. In this case the duals are
known to be type IIB superstring theory on $AdS_5\times S^5/\Gamma$ \cite{Kachru:1998ys}.

There is another class of conformal $\N=2$ theories based on the gauge
group $\Sp(N)$.\footnote{For us, $\Sp(N)$ has rank $N$, so $Sp(1)\equiv\SU(2)$ although
it is sometimes denoted $\Sp(2N)$ or even $\text{USp}(2N)$.} These
theories have an $2^{\rm nd}$-rank anti-symmetric tensor
and four fundamental hypermultiplets and are conformal
for any $N$. This class of theories is thought to be equivalent to
type IIB theory on $AdS_5\times S^5/\Z_2$, where the $\Z_2$ action
fixes an $S^3\subset S^5$, with four D7-branes wrapped around the $S^3$
and filling $AdS_5$ \cite{Fayyazuddin:1998fb,Aharony:1998xz}. 
What is particularly interesting about his case
is that, in common with the $\N=4$ theory,
there are detailed predictions for instanton contributions to
particular correlation functions \cite{Gutperle:1999xu} which merit investigation
from the gauge theory side. This paper sets up the formalism which
allows such a comparison. Near the completion of this paper, there
appeared a paper by Gava et al.~\cite{Gava:1999ky} which, as well as
setting up the same formalism leading to the large-$N$ instanton
measure, performs this comparison and finds perfect agreement with the
AdS/CFT correspondence.

Consider the gauge theory in more detail. It is useful to use a mock
$\N=4$ notation for the vector multiplet and anti-symmetric tensor
hypermultiplet. So the Weyl fermions are labelled $\lambda^A$,
$A=1,\ldots,4$, where $A=1,3$ correspond to the vector multiplet, and are
therefore adjoint-valued, while $A=2,4$ correspond to the
hypermultiplet. All fields can be thought of as $2N\times2N$
dimensional hermitian matrices subject to
\begin{equation}
X=\pm JX^TJ\ ,
\end{equation}
where $J$ is the usual symplectic matrix
\begin{equation}
J=\begin{pmatrix}0 & 1_{\sst[N]\times[N]}\\ -1_{\sst[N]\times[N]} & 0\end{pmatrix}\ ,
\end{equation}
and the upper (lower) sign corresponds to the vector multiplet (anti-symmetric
tensor hypermultiplet). With this restriction we see that the adjoint
representation of $Sp(N)$ has dimension $N(2N+1)$ while the
anti-symmetric tensor representation has dimension $N(2N-1)$.

In addition, we must add four fundamental hypermultiplets consisting
of eight $\N=1$ chiral superfields transforming with respect to an 
$O(8)$ flavour symmetry.
The theory has vanishing beta function for any $N$ and when $N=1$ the
only the fundamental hypermultiplets remain and the theory is
equivalent to the conformal $\SU(2)$ theory with four fundamental hypermultiplets.

\section{The Instanton Moduli Space and Measure}

The ADHM construction of instanton solutions \cite{ADHM} for gauge group $\Sp(N)$
is described in \cite{Corrigan:1978ce}. However, rather than use the
language of quaterions, we find it more in keeping with the string
theory connection to think of $\Sp(N)$ as embedded in $\SU(2N)$, and
use the ADHM construction of the latter suitably restricted. For a
review of the the ADHM construction for unitary groups see \cite{KMS,MO-III}.

The instanton solution at charge $k$ in $\SU(2N)$, is
described by an $(2N+2k)\times2k$ dimensional matrix $a$, and its
conjugate, with the form
\begin{equation}
a=\begin{pmatrix} w_\aD \\ a'_{\alpha\aD} \end{pmatrix}\ ,\qquad
\bar a=\begin{pmatrix} \bar w^\aD & \bar a^{\prime\aD\alpha} \end{pmatrix}\ ,
\end{equation}
where $w_\aD$ is a (spacetime) Weyl-spinor-valued $2N\times k$ matrix and
$a'_{\alpha\aD}$ is (spacetime) vector-valued $k\times k$ 
matrix. The conjugates are defined as
\begin{equation}
\bar w^\aD\equiv(w_\aD)^\dagger\ ,\qquad \bar
a^{\prime\aD\alpha}\equiv(a'_{\alpha\aD})^\dagger\ .
\end{equation}
The explicit $\SO(4)$ vector components of $a'$ are defined via
$a'_{\alpha\aD}=a'_n\sigma^n_{\alpha\aD}$ and the matrices $a'_n$ are
restricted to be hermitian: $(a'_n)^\dagger=a'_n$.

Up till now the discussion is valid for gauge group $\SU(2N)$,
however, in order to describe gauge group $\Sp(N)$ we only need
subject the variables to the additional conditions:
\begin{equation}
\bar w^\aD=
\epsilon^{\aD\bD}(w_\bD)^TJ\ ,\qquad 
(a'_{\alpha\aD})^T=a'_{\alpha\aD}\ ,
\end{equation}
The remaining $4k(N+k)$ variables are still an over parametrization of
the instanton moduli space which is obtained by a hyper-k\"ahler
quotient construction. One first imposes the ADHM constraints:
\begin{equation}
D^\aD_{\ \bD}\equiv\bar w^\aD w_\bD+\bar a^{\prime\aD\alpha}a'_{\alpha\bD}=\lambda
\delta^{\aD}_{\ \bD}\ ,
\label{badhm}\end{equation}
where $\lambda$ is an arbitrary constant. The ADHM moduli space is then
identified with the space of $a$'s subject to \eqref{badhm} modulo the
action of an $\O(k)$ symmetry which acts on the instanton indices of
the variables as follows
\begin{equation}
w_\aD\rightarrow w_\aD U\ ,\qquad a'_{\alpha\aD}\rightarrow
Ua'_{\alpha\aD}U^T\ ,\quad U\in\O(k).
\end{equation}
Since there are $3k(k-1)/2$ constraints \eqref{badhm} and $\O(k)$ has
dimension $k(k-1)/2$ the dimension of the ADHM moduli space is $4k(N+1)$.

The final piece of the story is the explicit construction of the
self-dual gauge field itself. To this end, we define the matrix 
\begin{equation}
\Delta(x)=\begin{pmatrix}w_\aD \\ x_{\alpha\aD}1_{\sst[k]\times[k]}+
a'_{\alpha\aD}\end{pmatrix}\ ,
\end{equation}
where $x_{\alpha\aD}$, or equivalently $x_n$, is a point in spacetime.
For generic $x$, the $(2N+2k)\times2N$ dimensional complex-valued matrix $\U(x)$
is a basis for ${\rm ker}(\bar\Delta)$:
\begin{equation}
\bar\Delta U= 0 = \bar U\Delta\ ,
\label{uan}\end{equation}
where $U$ is orthonormalized according to
\begin{equation}
\bar U U = \ 1_{{\sst[2N]}\times{\sst [2N]}}\,.
\label{udef}\end{equation}
The self-dual gauge field is then simply
\begin{equation}v_n = 
\bar U \partial_{n}U\ .
\label{vdef}\end{equation}
It is straightforward to show that $v_n$ is anti-hermitian and
satisfies $v_n=J(v_n)^TJ$ and so is valued in the Lie algebra of $\Sp(N)$.

The fermions in the background of the instanton lead to new Grassmann
collective coordinates. Associated to the fermions in the adjoint and
anti-symmetric tensor representations, $\lambda^A$, these collective
coordinates are described by the $(2N+2k)\times k$ matrices $\M^A$,
and their conjugates
\begin{equation}
\M^A=\begin{pmatrix} \mu^A \\ \M^{\prime A}_\alpha \end{pmatrix}\
,\qquad
\bar\M^A=\begin{pmatrix} \bar\mu^A & \bar\M^{\prime\alpha A} \end{pmatrix}\ ,
\end{equation}
where $\mu^A$ are $2N\times k$ matrices and $\M^{\prime A}_\alpha$ are Weyl-spinor-valued
$k\times k$ matrices. The conjugates are defined by
\begin{equation}
\bar\mu^A\equiv(\mu^A)^\dagger\ ,\qquad
\bar\M^{\prime\alpha A}\equiv(\M^{\prime A}_\alpha)^\dagger\ ,
\end{equation}
and the constraint $\bar\M^{\prime A}_\alpha=\M^{\prime A}_\alpha$ is
imposed. 

Up till now the construction of the fermions is valid for the gauge
group $\SU(2N)$. For $\Sp(N)$ all we need do is impose the additional conditions
\begin{equation}
\bar\mu^A=(-1)^{A+1}(\mu^A)^TJ\ ,\qquad
\qquad(\M^{\prime A}_\alpha)^T=(-1)^{A+1}\M^{\prime A}_\alpha\ .
\label{frest}\end{equation}
So, for instance, the matrices $\M^{\prime A}_\alpha$ are symmetric,
for the vector representation, and anti-symmetric, for the
anti-symmetric tensor representation. These symmetry properties can be
deduced from the ADHM tensor product construction in \cite{Corrigan:1979xi}.
Notice that the different 
conditions on the collective coordinates breaks the $\SU(4)$ symmetry
of the mock $\N=4$ notation to $\SU(2)\times\U(1)$, the true
$R$-symmetry of the $\N=2$ theory.
These fermionic collective coordinates are subject to analogues of the
ADHM constraints \eqref{badhm}:
\begin{equation}
\lambda^A_\aD\equiv\bar w_\aD\mu^A+\bar\mu^Aw_\aD+[\M^{\prime\alpha
A},a'_{\alpha\aD}]=0\ .
\label{fadhm}\end{equation}
One can check that there are consequently $2k(N+1)$ and $2k(N-1)$
fermionic collective coordinates corresponding to $\lambda^A$ for
the vector and anti-symmetric tensor representations, respectively. 
This is in accord with the index
theorem that counts the zero eigenvalue solutions to the Dirac
equation in the background of the instanton. The quantities
\begin{equation}
\xi_\alpha^A=k^{-1}{\rm tr}_k\,\M^{\prime A}_\alpha\ ,
\qquad \bar\eta^{\aD A}=k^{-1}{\rm tr}_k\,\bar w^\aD\mu^A\ ,
\end{equation}
in the vector multiplet, $A=1,3$, are the collective coordinates
corresponding to the four supersymmetric and four
superconformal zero modes.

Finally we have the fundamental hypermultiplets. They contribute
additional fermionic collective coordinates which can be described by
a $2N\times8$ dimensional matrix $\K$. These variables are not subject
to any additional ADHM-type constraints. There are no additional bosonic
collective coordinates associated to any of the scalar fields in the theory.

At lowest order in $g$, the instanton action, which is the pull-back
of the action to the instanton solution, is
\begin{equation}
S^k_{\text{inst}}={8\pi^2 k\over g^2}+S^k_{\text{quad}}\ ,
\end{equation}
where $S^k_{\text{quad}}$ is a certain four-fermion interaction which is
induced from the Yukawa interactions of the theory via tree-level
scalar exchange. This is where the $\N=4$ labelling comes in handy:
the interaction induced by the Yukawa couplings between the vector
multiplet and anti-symmetric tensor hypermultiplet
is precisely what would arise in the
$\N=4$ theory but with the variables restricted as in
\eqref{frest}. In addition, there
is a coupling between a bi-linear in the vector fermionic variables and the
hypermultiplet fermionic variables as in \cite{Dorey:1996bf} arising
form the Yukawa couplings between the vector multiplet and fundamental hypermultiplets:
\begin{equation}
S^k_{\text{quad}}={\pi^2\over
2g^2}{\rm tr}_k\big[\epsilon_{ABCD}\bar\M^A\M^B\BL^{-1}\bar\M^C\M^D-
\K\K^T\BL^{-1}(\bar\M^1\M^3-\bar\M^3\M^1)\big]\ .
\label{qact}\end{equation}
Here, $\BL$ is the following operator on $k\times k$
matrices:
\begin{equation}
\BL\cdot\Omega=\tfrac12\{\Omega, W^0\}+[a'_n,[a'_n,\Omega]]\ ,
\end{equation}
where $W^0\equiv\bar w^\aD w_\aD$.
Notice that the fundamental fermionic collective coordinates are only
coupled to the adjoint fermionic collective coordinates since there
are no Yukawa couplings between the former and the anti-symmetric 
fermionic collective coordinates.

Notice that the pattern of lifting of fermionic zero modes
in the fermion quadrilinear \eqref{qact} 
leads to a selection rule on the insertions of fermion modes. Suppose the insertions
in a correlator involve $n_{\rm adj}$ adjoint, $n_{\rm ast}$
antisymmetric tensor and $n_{\rm f}$ fundamental fermionic collective
coordinates, then for a non-trivial instanton contribution we need
\begin{equation}
n_{\rm adj}=n_{\rm ast}+n_{\rm f}\ ,
\end{equation}
subject to $n_{\rm adj}\leq4k(N+1)$, $n_{\rm ast}\leq4k(N-1)$ and
$n_{\rm f}\leq8k$. 
In particular, since the action does not lift the 8 exact supersymmetric and
superconformal zero-modes we have $n_{\rm adj}\geq8$.

Later we shall find it essential to bi-linearize this quadrilinear by
introducing bosonic variables $\chi_{AB}\equiv-\chi_{BA}$, six $k\times k$
matrices, subject to the reality condition
\begin{equation}
\bar\chi^{AB}\equiv(\chi_{AB})^\dagger=\tfrac12\epsilon^{ABCD}\chi_{CD}\ .
\end{equation}
We can write $\chi_{AB}$ as an explicit $\SO(6)$ vector $\chi_a$
with components
\begin{equation}\begin{split}
&\chi_1=\sqrt2(\chi_{12}+\chi_{34})\ ,\quad\chi_2=i\sqrt2(\chi_{12}-\chi_{34})\
,\quad\chi_3=\sqrt2(\chi_{14}+\chi_{23})\ ,\\
&\chi_4=i\sqrt2(\chi_{14}+\chi_{23})\ ,\quad\chi_5=i\sqrt2(\chi_{13}-\chi_{24})\
,\quad\chi_6=\sqrt2(\chi_{13}+\chi_{24})\ ,
\end{split}\end{equation}
normalized so that $\chi_a\chi_a\equiv\epsilon^{ABCD}\chi_{AB}\chi_{CD}$.
Taking account the following symmetry properties
\begin{equation}
(\bar\M^{[A}\M^{B]})^T=(-1)^{A+B+1}\bar\M^{[A}\M^{B]}\ ,\qquad
(\K\K^T)^T=-\K\K^T\ ,
\end{equation}
the auxiliary variables $\chi_{AB}$ are subject in addition to
\begin{equation}
\chi_{AB}=(-1)^{A+B+1}(\chi_{AB})^T\ .
\end{equation}
So $\chi_a$, $a=1,\ldots,4$, are symmetric and $\chi_a$, $a=5,6$, are anti-symmetric.
These conditions obviously only respect a
$\SU(2)\times\U(1)$ subgroup of $\SU(4)$, the $R$-symmetry
of the $\N=2$ theory. The transformation we want is then 
\begin{equation}
e^{-S^k_{\text{quad}}}=(\text{det}_s\BL)^2(\text{det}_a\BL)
\int d\chi\,\exp\big[-\epsilon^{ABCD}{\rm tr}_k\,\chi_{AB}\BL\chi_{CD}
+4\pi ig^{-1}{\rm tr}_k\,\chi_{AB}\Lambda^{AB}\big]\ ,
\label{aact}\end{equation}
where $\text{det}_s\BL$ and $\text{det}_{a}\BL$ are the determinants
of $\BL$ evaluated on a
basis of symmetric and anti-symmetric matrices, respectively, and we
have defined the fermion bi-linear
\begin{equation}
\Lambda^{AB}={1\over2\sqrt2}\bar\M^{A}\M^{B}-
{1\over4\sqrt2}\delta^{A2}\delta^{B4}\K\K^T\ .
\end{equation}
The determinant factors in \eqref{aact} are conveniently cancelled by
factors that arise from integrating-out pseudo collective coordinates
corresponding to the scalars from the vector and anti-symmetric tensor
multiplets (see \cite{MO-III}).

The measure on the ADHM moduli space can be deduced from \cite{meas1,meas2,MO-III}
and is simply the `flat' Cartesian measure for all the ADHM variables,
including the auxiliary variables $\chi$, with the ADHM constraints \eqref{badhm} and
\eqref{fadhm} imposed by explicit delta functions, weighted by the
exponential of the action on the right-hand side of \eqref{aact}. 
Up to an overall normalization factor
\begin{equation}\begin{split}
\int d\mu^k_{\rm phys}\,e^{-S^k_{\rm inst}}&
={1\over\text{Vol}\,\O(k)}\int da'\,dw\,d\chi\,d\M'\,d\mu\,d\K\,\\
&\times \delta(D^\aD_{\ \bD})\,\delta(\lambda^A_\aD)
\exp\big[-\epsilon^{ABCD}{\rm tr}_k\,\chi_{AB}\BL\chi_{CD}
+4\pi ig^{-1}{\rm tr}_k\,\chi_{AB}\Lambda^{AB}\big]\ ,
\label{meas}\end{split}\end{equation}
where the delta functions impose the
bosonic and fermionic ADHM constraints \eqref{badhm} and \eqref{fadhm}.
As in the $\N=4$ case, this measure has the natural interpretation in
terms of branes. First of all the $\N=2$ gauge theory can be described
by the low energy limit of a set of $N$ D3-branes moving tangent to
of one of the four orientifold O7-planes \cite{Banks:1996nj,Douglas:1997js} in the type
IIB orientifold of Sen \cite{Sen:1996vd}. In this construction there are four D7-branes
on top of each of the O7-planes. The gauge theory on the world volume
of the D3-branes is precisely our $\N=2$
supersymmetric $\Sp(N)$ gauge theory, with an anti-symmetric tensor
hypermultiplet and four fundamental hypermultiplets. The latter fields
arise from strings stretched between the D3- and the four D7-branes.

Now on top of this construction
we include $k$ D-instantons. The matrix theory on the `world volume'
of the D-instantons will reproduce the
ADHM construction and measure described above. In order to see this,
it is useful to start from D instantons is flat ten-dimensional space
which are described by the dimensional reduction of $\N=1$
supersymmetric gauge theory in ten dimensions. The Lorentz group in
ten-dimensional Euclidean space is broken to
$H=\SO(4)_1\times\SO(4)_2\times\SO(2)\subset\SO(10)$, where the first
factor corresponds to the directions along the D3-brane world-volume,
the second factor to the directions in the D7/O7 world-volume 
orthogonal to the D3-branes and finally the last factor corresponds to
the two directions orthogonal to the D7/O7 world volume.
The components of ten-dimensional gauge field decompose as 
\begin{equation}
{\bf10}\rightarrow({\bf4},{\bf1},{\bf1})^s\oplus({\bf1},{\bf4},{\bf1})^s
\oplus({\bf1},{\bf1},{\bf2})^a
\label{dvec}\end{equation}
corresponding in the ADHM construction 
to $a'_n$, $\chi_a$, $a=1,\ldots,4$ and $\chi_a$,
$a=5,6$, respectively. The superscripts in \eqref{dvec} label the
projections on the $\U(2N)$ valued variables
imposed in the orientifold background, where $s$ means symmetric and $a$
means anti-symmetric.

In order to describe the fermions we consider
the covering group of $H$, $\bar H=\SU(2)_{L_1}\times\SU(2)_{R_1}\times
\SU(2)_{L_2}\times\SU(2)_{R_2}\times\U(1)$. The Majorana-Weyl fermion
of the ten-dimensional theory decomposes as
\begin{equation}
{\bf16}\rightarrow({\bf2},{\bf1},{\bf2},{\bf1})^s_1
\oplus({\bf2},{\bf1},{\bf1},{\bf2})^{a}_1\oplus({\bf1},{\bf2},{\bf2},{\bf1})^s_{-1}\oplus
({\bf1},{\bf2},{\bf1},{\bf2})^{a}_{-1}
\end{equation}
under $\bar H$. Again the superscripts indicate the projection imposed
by the orientifold background. The first two components are identified with the 
ADHM variables $\M^{\prime A}_\alpha$, for $A=1,3$ and $2,4$, respectively,
and the latter two with 
some additional variables $\lambda^\aD_A$ whose r\^ole will emerge
shortly. Notice that the subgroup $\SU(2)_{L_2}\times\U(1)\subset\bar
H$ is identified with the $R$-symmetry of the original $\N=2$ gauge theory.

The remaining ADHM variables correspond to the presence of the D3- and
D7-branes. For instance $w_\aD$ and $\mu^A$ are associated to strings stretched
between the D-instantons and the D3-branes and finally the variables $\K$ are
associated to strings stretched between the D-instantons and D7-branes
(there are no bosonic variables associated to these strings). We shall not
write down the full action for this theory, but it is a simple
generalization to include $\K$ of that for the $\N=4$ theory written
down in \cite{MO-III}. However, as explained in \cite{MO-III} it is useful to
introduce additional bosonic auxiliary $k\times k$ matrices $D_{\ \bD}^\aD$
transforming in a triplet of $\SU(2)_{R_1}$, and in the present
context anti-symmetric, whose significance will
emerge shortly. Now in the decoupling limit $\alpha'\rightarrow0$, the
coupling constant of the D-instanton `world volume' matrix theory 
goes to infinity which effectively
removes certain terms from the action. It is straightforward to show
that the remaining action is precisely that on the right-hand side of
\eqref{meas} and the variables $D_{\ \bD}^\aD$ and $\lambda^\aD_A$ act as Lagrange
multipliers that impose the bosonic and fermionic ADHM
constraints \eqref{badhm} and \eqref{fadhm}, respectively.

\section{The Large-$N$ Measure}

We now want to take find an approximation of the ADHM measure valid
in the large $N$-limit. The procedure is by now well documented
\cite{MO-III,N=2,orbi}. 

First of all it is advantageous to write the measure in terms of a set of
gauge invariant variables defined in terms of a $2k\times 2k$
dimensional matrix 
\begin{equation}
W^\aD_{\ \bD}=\bar w^\aD w_\bD\ .
\end{equation}
From this we define the following four $k\times k$ matrices
\begin{equation}
W^0=\bar w^\aD w_\aD\ ,\qquad W^c=(\tau^c)^\bD_{\ \aD}\bar w^\aD w_\bD\ .
\end{equation}
The matrix $W^0$ is symmetric while the three matrices
$W^c$ are anti-symmetric.
We now change variables from $w_\aD$ to the gauge invariant variables
above and the gauge transformations acting on the solution which
parameterize a coset. Since the measure is used to integrate gauge
invariant quantities we integrate over the gauge coset which yields a
volume factor. Up to numerical factors\footnote{For later use, the
numerical pre-factor goes like $N^{-2kN+k^2-k/2}$ at large $N$.} 
and for $N\geq k$ we have 
\begin{equation}
\int_{\text{coset}}dw\sim 
\int(\text{det}_{2k}W)^{N-k^2/2+1/4}\,dW^0\prod_{c=1,2,3}
dW^c\ .
\label{wcv}\end{equation}
In terms of the gauge invariant coordinates the ADHM constraints
\eqref{badhm} are linear:
\begin{equation}
W^c=i\bar\eta^c_{nm}[a'_n,a'_m]\ ,
\end{equation}
where $\bar\eta^c_{nm}$ is a 't Hooft eta-symbol defined in
\cite{MO-III}, and hence can be trivially integrated out.

On the fermionic side we need to integrate out the superpartners of the
gauge degrees-of-freedom. To isolate these variables we expand
\begin{equation}
\mu^A=w_{\aD}\zeta^{\aD A}+w_\aD\sigma^{\aD A}+\nu^A\ ,
\label{E44}\end{equation}
where the $\nu^A$ modes---the ones we need to integrate out---form a
basis for the kernel of $\bar w^\aD$, i.e.~$\bar w^\aD\nu^A=0$, 
The remaining variables are chosen to have the symmetry properties
\begin{equation}
(\zeta^{\aD A})^T=(-1)^{A+1}\zeta^{\aD A}\ ,\qquad
(\sigma^{\aD A})^T=(-1)^A\sigma^{\aD A}\ .
\end{equation}
Notice that the $\nu^A$ variables decouple from the fermionic
ADHM constraints \eqref{fadhm}. The change of variables from $\mu^A$ to
$\{\zeta^A,\sigma^A,\nu^A\}$ involves a Jacobian:
\begin{equation}
\int\, \prod_{A=1,2,3,4}d\mu^A=\int \left(\text{det}_{2k}W\right)^{-2k}\,\prod_{A=1,2,3,4}
d\zeta^A\,d\sigma^A\,d\nu^A\ . 
\end{equation}
We can now integrate out the $\nu^A$ variables:
\begin{equation}
\int\, \prod_{A=1,2,3,4} d\nu^A\, 
\exp\big[\sqrt8\pi ig^{-1}
{\rm tr}_k\,\chi_{AB}\bar\nu^A\nu^B\big]=2^{6k(N-k)}
(\pi/g)^{4k(N-k)}
\left({\rm det}_{4k}\chi\right)^{N-k}\ .
\label{E55}\end{equation}
Furthermore, the fermionic ADHM constraints \eqref{fadhm} 
can then be used to integrate-out the variables $\sigma^{\aD A}$.
The resulting expression is rather cumbersome and is, in any case, not
needed at this stage. Fortunately, in the large $N$ limit it
simplifies considerably.

We now proceed to take a large-$N$ limit of the measure using a
steepest descent method. As explained
in \cite{MO-III} it is useful to scale $\chi_{AB}\rightarrow\sqrt
N\chi_{AB}$ because this exposes the three terms which contribute to the
large-$N$ `saddle-point' action:
\begin{equation}S_{\rm
eff}=-\log{\rm det}_{2k}W-\log{\rm det}_{4k}\chi+{\rm
tr}_K\,\chi_a\BL\chi_a\,.
\label{E57}\end{equation} 
The first and second terms come from \eqref{wcv} and \eqref{E55} while the
final term comes from \eqref{meas}.

We can now perform a steepest descent approximation of the measure
which involves finding the minima of the effective action with
respect to the variables $\chi_{AB}$, $W^0$ and $a'_n$. The resulting
saddle-point equations are\footnote{As usual we shall frequently
swap between the two representations $\chi_a$ and $\chi_{AB}$ for $\SO(6)$ vectors.}
\begin{equation}
\begin{split}
\epsilon^{ABCD}\left(\BL\cdot\chi_{AB}\right) \chi_{CE}\ &=\
\tfrac12\delta^D_E \,1_{\sst [K]\times[K]}\ ,\\
\chi_a\chi_a\ &=\ \tfrac12(W^{-1})^0\ ,\\
[\chi_a,[\chi_a,a'_n]]\ &=\ i\bar\eta^c_{nm}[a'_m,(W^{-1})^c]\ .
\label{E60}
\end{split}
\end{equation}
where we have introduced the matrices
\begin{equation}
(W^{-1})^0=
{\rm tr}_2\,W^{-1},\qquad
(W^{-1})^c=
{\rm tr}_2\,\tau^cW^{-1}\ .
\label{E60.1}\end{equation}

We note that the expression for the effective action \eqref{E57}
and the saddle point equations \eqref{E60}
look identical to those derived in \cite{MO-III}
for the $\N=4$ theory, however, the difference is the symmetry
properties of the variables $\{a'_n,W^0,\chi_a\}$ which will have to
be taken into account.
As in the $\N=4$ case we look for a solution with $W^c=0$,
$c=1,2,3$, which means that the instantons are embedded in mutually commuting
$\SU(2)$ subgroups of the gauge group. In this case 
equations \eqref{E60} are equivalent to
\begin{equation}
[a'_n,a'_m]=[a'_n,\chi_a]=[\chi_a,\chi_b]=0\ ,\quad
W^0=\tfrac12(\chi_a\chi_a)^{-1}\ .\label{echo}
\end{equation}
The final equation can be viewed as giving the value of $W^0$, whose
eigenvalues are the
instanton sizes at the saddle-point. Clearly $\chi_a\chi_a$ and
$W^0$ must be non-degenerate.

The solution space
to these equations (up to the auxiliary $\O(k)$ symmetry) 
is composed of distinct branches. On the first branch
\begin{equation}\begin{split}
a'_n&={\rm diag}\big(\,-\com^1_n\,,\,\ldots\,,\,-\com^k_n\,\big)\ ,
\\
\chi_a&=\begin{cases}{\rm
diag}\big(\rho_1^{-1}\hat\Theta_a^{1}\,,\,\ldots\,,\,\rho_k^{-1}\hat\Theta_a^k\big)
& a=1,\ldots,4\ ,\\ 0 & a=5,6\ .\end{cases}
\label{firb}
\end{split}\end{equation}
where $\hat\Theta_a$ is a unit four-vector, i.e.~$\hat\Theta_a\equiv0$ for
$a=5,6$. A second branch of solutions only exists for even instanton
number $2k$ and is, in block form,
\begin{equation}\begin{split}
a'_n&={\rm diag}\big(\,-\com^1_n\,,\,\ldots\,,\,-\com^k_n\,\big)
\otimes1_{\sst[2]\times[2]}\ ,\\
\chi_a&=
{\rm diag}\big(\rho_1^{-1}\sfc_a^{1}\,,\,\ldots\,,\,\rho_k^{-1}\sfc_a^k\big)\otimes
\begin{cases}1_{\sst[2]\times[2]} & a=1,\ldots,4\ ,\\ \tau^2 & a=5,6\ ,\end{cases} 
\label{secb}
\end{split}\end{equation}
where $\sfc_a$ is a unit six-vector. Notice that the sign of $\sfc_a^i$,
$a=5,6$, can be reversed by an $\O(2k)$ transformation $\tau^1$. So
physically $\sfc_a^i$ is coordinate on the orbifold $S^5/{\mathbb Z}_2$.
In these solutions $X_n^i$ are the
position of the individual instantons in ${\mathbb R}^4$ and $\rho_i$
are their scale sizes. 

The first branch can be interpreted as parameterizing, via
$\{X^i_n,\rho_i,\hat\Theta_a^i\}$, the position of
$k$ objects, the  D-instantons, moving in $AdS_5\times S^3$,
where the $S^3$ is precisely the orbifold singularity of $S^5/{\mathbb
Z}_2$ of the dual type IIB superstring background. Since these
instantons are confined to the singularity they are
fractional \cite{Douglas:1996sw,Douglas:1997xg,Diaconescu:1998br}. The second branch
describes how the fractional D-instantons confined to the singularity on the first
branch can move off in pairs to explore the whole of $S^5/{\mathbb Z}_2$ parametrized
by $\sfc_a^i$. Of course there are also mixed branches consisting of
fractional instantons on, and `normal' instanton off, the
singularity. However, we shall not need these more general
solutions in our analysis.

In principle, in order to do a saddle-point analysis 
we have to expand the effective action 
around these general solutions to sufficient order 
to ensure that the fluctuation integrals converge. In
general because the Gaussian form has zeros whenever two D-instantons
coincide one has to go to quartic order in the fluctuations. 
Fortunately, as explained in \cite{MO-III}, 
we do not need to expand about these most general solution to the
saddle-point equations to quartic order since this is equivalent to expanding
to the same order around the most degenerate solution where all the 
D-instantons are at the same point.
The resulting quartic action has flat directions corresponding to the 
relative positions of the D-instantons. However, when the fermionic
integrals are taken into account the integrals over these relative
positions turn out to be convergent and hence these degrees-of-freedom
should be viewed as fluctuations around rather than facets of the maximally
degenerate solution. The variables left un-integrated, since they are
not convergent, are the centre-of-mass coordinates.
In the present situation, as in the orbifold
theories \cite{orbi}, there is an important further implication: the
number of centre-of-mass coordinates (corresponding to the
non-damped integrals) can depend on exactly what fermionic insertions are
made into measure since this affects the convergence properties of the
bosonic integrals. We will see that this is closely connected with the
different branches of the solution space.

For the first branch of solutions, the maximally degenerate solution 
is \eqref{firb} with all the
instantons at the same point:
\begin{equation}\begin{split}
a'_n&=-X_n1_{\sst[k]\times[k]}\ ,\\
\chi_a&=\begin{cases}
\rho^{-1}\hat\Theta_a1_{\sst[k]\times[k]} & a=1,\ldots,4\ ,\\
0 & a=5,6\ .\end{cases}
\label{fmaxs}\end{split}
\end{equation}
For the second branch with instanton number $2k$, the maximally
degenerate solution is \eqref{secb} with all the instantons at the
same point:
\begin{equation}\begin{split}
a'_n&=-X_n1_{\sst[2k]\times[2k]}\ ,\\
\chi_a&=\rho^{-1}\sfc_a1_{\sst[k]\times[k]}\otimes\begin{cases}
1_{\sst[2]\times[2]} & a=1,\ldots,4\ ,\\
\tau^2 & a=5,6\ .\end{cases}
\label{smaxs}\end{split}
\end{equation}
In \eqref{fmaxs} $\hat\Theta_a$ parametrizes $S^3$, whereas in
\eqref{smaxs}, $\sfc_a$ parametrizes $S^5/{\mathbb Z}_2$.

The expansion of the effective action to quartic order around these solution 
can be deduced from the analysis of \cite{MO-III}, although the first
case describing the fractional D-instantons is somewhat simpler and we
describe it first. After
integrating-out the fluctuations in the variables $W^0$ which are lifted at Gaussian
order, as in \cite{MO-III}, the quartic fluctuations are governed by the
action which looks identical to the action of ten-dimensional $\N=1$
supersymmetric $\U(k)$ gauge theory dimensionally reduced to zero dimensions:
\begin{equation}
S_{\rm b}=-{1\over2}{\rm tr}_k\,\big[A_\mu,A_\nu\big]^2\ ,
\end{equation}
where the gauge field has components
\begin{equation}
A_\mu=N^{1/4}\big(\rho^{-1}a'_n,\rho\chi_a\big)\ .
\end{equation}
However, the components of the gauge field are subject to
\begin{equation}
(A_\mu)^T=A_\mu\ ,\quad\text{for}\ \mu=1,\ldots,8\ ,\quad
(A_\mu)^T=-A_\mu\ ,\quad\text{for}\ \mu=9,10\ .
\end{equation}
In other words, the matrix model only has $\O(k)\subset\U(k)$ symmetry.
The centre-of-mass parameters of the maximally degenerate solution,
$X_n$ and $\rho^{-1}\hat\Theta_a$, correspond to the trace parts
of $A_\mu$, $\mu=1,\ldots,8$, and these decouple from the action.

Now we turn to the fermionic sector which are coupled to the
bosonic variables in \eqref{aact}. First of all we fulfill our
promise to deal with the fermionic ADHM constraints. To leading order
in $1/N$, these constraints read
\begin{equation}
2\rho^2\sigma_\aD^A=-\tfrac12[\delta W^0,\zeta_\aD^A]-[\M^{\prime\alpha
A},a'_{\alpha\aD}]\ .
\label{lnfadhm}\end{equation}
So the integrals over the $\sigma^{\aD A}$ variables soak up the
delta-functions imposing the fermionic ADHM constraints, as promised.
In \eqref{lnfadhm}, $\delta W^0$ are the fluctuations in $W^0$, all of which are
lifted at Gaussian order. Due to a cross term we can effectively replace
$\delta W^0$ with $-4\rho^3\hat\Theta\cdot\chi$ at leading order (see
\cite{MO-III}). Collecting all the leading order terms, the fermion couplings
are
\begin{equation}\begin{split}
S_{\rm f}=i\Big({8\pi^2 N\over g^2}\Big)^{1/2}
{\rm tr}_k\Big[&
-2\rho^2(\hat\Theta\cdot\chi)\hat\Theta_{AB}
\zeta^{\aD A}\zeta_\aD^B+\rho^{-1}\hat\Theta_{AB}\big[a'_{\alpha\aD},
{\cal M}^{\prime\alpha A}\big]\zeta^{\aD B}\\
&\qquad\qquad\qquad+\chi_{AB}\big(\rho^2 
\zeta^{\aD A}\zeta_\aD^B+{\cal M}^{\prime
\alpha A}{\cal M}^{\prime B}_\alpha\big)-\tfrac12\chi_{24}\K\K^T
\Big] \ .
\label{fermc}\end{split}\end{equation}
The fermionic variables $\M^{\prime A}_\alpha$ and 
$\zeta^{\aD A}$ can be amassed into two $\SO(8)$ spinors $\Psi$ and
$\Phi$:
\begin{equation}\begin{split}
\Psi&=\sqrt{\pi\over2 g}N^{1/8}\big(\rho^{-1/2}\M^{\prime 1}_\alpha,
\rho^{-1/2}\M^{\prime
3}_\alpha,\rho^{1/2}\zeta^{\aD1},\rho^{1/2}\zeta^{\aD3}\big)\ ,\\
\Phi&=\sqrt{\pi\over2 g}N^{1/8}\big(\rho^{-1/2}\M^{\prime 2}_\alpha,
\rho^{-1/2}\M^{\prime
4}_\alpha,\rho^{1/2}\zeta^{\aD2},\rho^{1/2}\zeta^{\aD4}\big)\ ,
\end{split}\end{equation}
transforming as an ${\bf8}_s$ and ${\bf8}_{\bar s}$, respectively, and
as $k\times k$ matrices subject to
\begin{equation}
\Psi^T=\Psi\ ,\qquad\Phi^T=-\Phi\ .
\end{equation}
We also define rescaled fundamental collective coordinates
\begin{equation}
\K\rightarrow\sqrt{4g\over\pi}N^{-1/8}\K\ .
\end{equation}
In terms of the new variables, the fermion couplings can be written in
an elegant way as
\begin{equation}
S_{\rm f}=i\,{\rm
tr}_k\big(\Psi\Gamma_{\hat\mu}[A_{\hat\mu},\Phi]+\Psi[A_9-iA_{10},\Psi]
+\Phi[A_9+iA_{10},\Phi]-(A_9+iA_{10})\K\K^T\big)\ ,
\label{fc}\end{equation}
with the appropriate $\SO(8)$ inner products between the spinors.
Here $\Gamma_{\hat\mu}$, $\hat\mu=1,\ldots,8$ is a representation of the
$D=8$ Clifford algebra which depends upon $\hat\Theta$.\footnote{The
construction of this Clifford algebra can be deduced from the similar
construction in $D=10$ for the $\N=4$ theory in \cite{MO-III}.}

The couplings \eqref{fc} do not involve the trace parts of
$\Psi$ corresponding to the eight supersymmetric and superconformal
zero-modes of the instanton, which we denoted previously as
$\xi^A_\aD$ and $\bar\eta^{\aD A}$, for $A=1,3$.
Separating out the integrals over these variables and the bosonic
centre-of-mass variables, the measure at large $N$ has the
form\footnote{We have kept track of power of $N$ and $g$ but not other
numerical factors.} 
\begin{equation}
\int d\mu^ke^{-S^k_{\rm inst}}\Big\vert_{1^{\rm st}\text{ branch}}\ 
\underset{N\rightarrow\infty}=\ 
g^4Ne^{2\pi ik\tau}\int
{d^4X\,d\rho\over\rho^5}\,d^3\hat\Theta\,\prod_{A=1,3}d^2\xi^A\,d^2\bar\eta^A\cdot
{\cal Z}_{\O(k)}\ ,
\label{lnm}\end{equation}
where ${\cal Z}_{\O(k)}$ is the partition function of an $\O(k)$
matrix model:
\begin{equation}
{\cal Z}_{\O(k)}={1\over{\rm Vol}\,\O(k)}\int d\hat A\,d\hat\Psi
\,d\Phi\,d\K\,e^{-S(A_\mu,\Psi,\Phi,\K)}\ ,
\end{equation}
where the hat indicates the traceless parts and
$S(A_\mu,\Psi,\Phi,\K)=S_{\rm b}+S_{\rm f}$. This matrix model is
identical to the one written down in \cite{Gutperle:1999xu} and describes the
dynamics of $k$ D-instantons in flat space. In other words, at large
$N$ the D3-branes have disappeared explicitly in the description of
D-instanton dynamics and their only effect is to change the 
centre-of-mass measure from ${\mathbb R}^{10}$ to $AdS_5\times S^3$, along with the 8
supersymmetric and superconformal fermionic integrals which are
associated to the 8 symmetries broken by the instanton. One can show
that the integrals over all the bosonic variables in the definition 
of ${\cal Z}_{\O(k)}$ are actually convergent so that ${\cal
Z}_{\O(k)}$ is some finite numerical factor (this is important for
getting the $N$ counting of the answer correct).

Now we turn to the expansion around the maximally degenerate solution
for the second branch of solutions \eqref{secb} with instanton number
$2k$. In this case the solution is not proportional to the identity
and many of the fluctuations, those which do not commute $\chi_a$, for
$a=5,6$, are lifted at Gaussian order. In fact it is not difficult to
see that the resulting leading order term can be deduced from the
$\O(2k)$ matrix model constructed above by expanding around
$\chi_a=\rho^{-1}\sfc_a1_{\sst[k]\times[k]}\otimes\tau^2$, $a=5,6$.  
The solution only commutes with a subgroup $\U(k)\subset\O(2k)$, and
so the resulting model only has $\U(k)$ symmetry. After a suitable
gauge fixing (described in the context of the orbifold models in 
\cite{orbi}) and after integrating-out the Gaussian
fluctuations, which include all the fundamental collective coordinates
$\K$, one is left with the partition function of 
a $\U(k)$ matrix model which is
identical to that which appears in the $\N=4$ calculation namely 
ten-dimensional $\N=1$ gauge theory dimensionally reduced to zero
dimensions. The $\U(k)$ gauge field $A'_\mu$ is embedded in the $\O(2k)$
model variables $A_\mu$ as
\begin{equation}
A_\mu=A'_\mu\otimes\begin{cases}1_{\sst[2]\times[2]} & \mu=1,\ldots,8\
,\\ \tau^2 & \mu=9,10\ .\end{cases}\ .
\end{equation}
with a similar relation for the $\U(k)$ fermions:
\begin{equation}
\Psi=\Psi'\otimes1_{\sst[2]\times[2]}\ ,\qquad\Phi=\Phi'\otimes
\tau^2\ .
\end{equation}
The 16 fermions $\Psi'$ and $\Phi'$ combine into the 16 component
Majorana Weyl fermion of the ten-dimensional theory.
The trace parts of the gauge field $A'_\mu$ 
separate out to give an integral over $AdS_5\times S^5/{\mathbb Z}_2$ along
with 16 fermionic integrals which include the 8 supersymmetric and superconformal
zero modes along with the 8 components of $\Phi$ proportional to
$1_{\sst[k]\times[k]}\otimes\tau^2$, which for conformity of
notation we denote $\xi_\alpha^A$
and $\bar\eta^{\aD A}$, for $A=2,4$. The final expression is
\begin{equation}
\int d\mu^{2k}_{\rm phys}e^{-S^{2k}_{\rm inst}}\Big\vert_{2^{\rm nd}
\text{ branch}}\ \underset{N\rightarrow\infty}=\  
g^8\sqrt Ne^{4\pi
ik\tau}\int{d^4X\,d\rho\over\rho^5}\,d^5\sfc\,\prod_{A=1,\ldots,4}\,
d^2\xi^A\,d^2\bar\eta^A\cdot{\cal Z}_{\SU(k)}\ ,
\end{equation}
which is identical to the measure in the $\N=4$ case. Again, it is
important that bosonic integrals in the definition of the partition
function ${\cal Z}_{\SU(k)}$ are all convergent and this factor is
some overall numerical factor. Our analysis
makes it clear that the second branch is actually contained within the
more general first branch. The key point is that the convergence of
the integrals of the bosonic variables in the $\O(2k)$ matrix model
depends on the `fermionic context', i.e.~on what fermion insertions
are made 
into the partition function. The second branch corresponds to inserting
the 8 modes $\xi_\alpha^A$ and $\bar\eta^{\aD A}$, $A=2,4$, into the
$\O(2k)$ partition function. The integrals over
the components of $A_\mu$, for $\mu=9,10$, proportional to
$1_{\sst[k]\times[k]}\otimes\tau^2$, are not damped after taking
into account the now smaller number of fermionic integrals. These divergent integrals
need to be separated out as additional centre-of-mass coordinates
leading to an integral over $S^5/{\mathbb Z}_2$ rather
than $S^3$.

\acknowledgments

I would like to thank Michael Gutperle for very helpful conversations.

Note: near the completion of this work
there appeared a paper by Gava et al.~\cite{Gava:1999ky}. That paper
has identical conclusions to the present work but goes much further and
provides a very detailed comparison between instanton contributions to
certain correlators in the gauge theory and the string theory. The
results are in perfect agreement with the AdS/CFT correspondence.

\end{document}